# Some Unique Constants Associated with Extremal Black Holes

*C Sivaram*

Indian Institute of Astrophysics, Bangalore, 560 034, India

Telephone: +91-80-2553 0672; Fax: +91-80-2553 4043

e-mail: sivaram@iiap.res.in

*Kenath Arun*

Christ Junior College, Bangalore, 560 029, India

Telephone: +91-80-4012 9292; Fax: +91-80- 4012 9222

e-mail: kenath.arun@cjc.christcollege.edu

**Abstract:** In recent papers we had developed a unified picture of black hole entropy and curvature which was shown to lead to Hawking radiation. It was shown that for any black hole mass, holography implies a phase space of just one quantum associated with the interior of the black hole. Here we study extremal rotating and charged black holes and obtain unique values for ratios of angular momentum to entropy, charge to entropy, etc. It turns out that these ratios can be expressed in terms of fundamental constants in nature, having analogies with other physical systems, like in condensed matter physics.

In recent papers (Sivaram & Arun, 2010; 2011) we had developed a unified picture of black hole entropy and curvature which was shown to lead to Hawking radiation. Invariant underlying relations connecting curvature with Hawking luminosity and temperature were studied. It was also shown that irrespective of the mass of the black hole, holography implies a phase space associated with the interior volume which remarkably turned out to be just one quantum of phase space filled with just one photon: (Sivaram & Arun, 2009)

$$\left(d^3 p\, d^3 x\right)_{\text{inside horizon}} = \left(\frac{\hbar c^2}{GM}\right)^3 \times \left(\frac{GM}{c^2}\right)^3 = \hbar^3 \qquad \ldots (1)$$

Here we study extremal rotating and charged black holes and obtain unique values (expressed only in fundamental physical constants!) for ratios of angular momentum to entropy, charge to entropy, etc. These relations have analogues in familiar physical systems, for instance in condensed matter physics.

In earlier papers, we had obtained for any black hole: Entropy × horizon curvature = $k_B \frac{c^3}{\hbar G}$, the Planck curvature = $\frac{c^3}{\hbar G}$, and similar relations for other associated quantities.

Now consider extreme rotating black holes whose angular momentum (for any mass *M*) is given as:

$$J = \frac{GM^2}{c} \qquad \ldots (2)$$

The area of the horizon is:

$$A \sim \frac{G^2 M^2}{c^4} \qquad \ldots (3)$$

So that the angular momentum per unit area is:

$$\sigma = \frac{J}{A} = \frac{c^3}{G} \qquad \ldots (4)$$

Now the entropy per unit area, works out as:

$$S = k_B \frac{GM^2}{\hbar c} \bigg/ \frac{G^2 M^2}{c^4} \approx k_B \frac{c^3}{\hbar G} \qquad \ldots (5)$$

(where the black hole entropy is $\sim k_B \frac{GM^2}{\hbar c}$)

Equations (4) and (5), imply the remarkable relation that for extremal black holes (of any mass), the ratio of angular momentum to entropy, i.e.:

$$\frac{J}{S} = \frac{\hbar}{k_B} \qquad \ldots (6)$$

This is just the same as for an elementary particle or quantum of radiation! (spin $\hbar$, entropy $k_B$).

Thus the black hole interior volume (for whatever black hole mass in any number of spatial dimensions) was earlier shown to be associated with a phase space of just $\hbar^3$ (having a single photon). Now the ratio of angular momentum to entropy is just $\hbar/k_B$.

Also for extremely charged black holes: $Q^2 = GM^2$ $\qquad \ldots (7)$

This implies that the ratio of charge to angular momentum for extremal black holes (of any mass) is just:

$$\frac{Q^2}{J} = c \qquad \ldots (8)$$

Or a dimensionless 'black hole fine structure' constant of:

$$\frac{Q^2}{Jc} = 1 \qquad \ldots (9)$$

(analogous to $\alpha = \frac{e^2}{\hbar c}$ or $\frac{g^2}{\hbar c} \sim 1$ for strong interaction!)

Again the universal relation, $\frac{J}{Q^2} = \frac{1}{c}$ is similar to the von-Klitzing constant $\frac{\hbar}{e^2}$ in the quantum Hall effect or the Josephson constant $\frac{e^2}{\hbar}$, in superconductors (it was earlier indicated that black holes are diamagnetic, excluding flux, like superconductors).

Again the charge to entropy ratio is:

$$\frac{Q^2}{S} = \frac{\hbar c}{k_B} \qquad \ldots (10)$$

All the above relations involve only the fundamental constants, $\hbar, c, k_B$ and are even independent of G, although black holes are strongly gravitating systems!

Several similar properties of black holes corresponding to those of elementary particles including the fact that the gyromagnetic ratio of Kerr-Newman extremal black holes is 2, just the same as a Dirac particle were indicated earlier (Sivaram & Sinha, 1977).